\documentstyle[psfig]{mn}

\newif\ifAMStwofonts

\ifoldfss
  \ifCUPmtlplainloaded \else
    \NewTextAlphabet{textbfit} {cmbxti10} {}
    \NewTextAlphabet{textbfss} {cmssbx10} {}
    \NewMathAlphabet{mathbfit} {cmbxti10} {} 
    \NewMathAlphabet{mathbfss} {cmssbx10} {} 
  \fi
  \ifAMStwofonts
    \ifCUPmtlplainloaded \else
      \NewSymbolFont{upmath} {eurm10}
      \NewSymbolFont{AMSa} {msam10}
      \NewMathSymbol{\upi}     {0}{upmath}{19}
      \NewMathSymbol{\umu}     {0}{upmath}{16}
      \NewMathSymbol{\upartial}{0}{upmath}{40}
      \NewMathSymbol{\leqslant}{3}{AMSa}{36}
      \NewMathSymbol{\geqslant}{3}{AMSa}{3E}

      \let\leq=\leqslant 
      \let\geq=\geqslant 
    \fi
  \fi
\fi 

\ifnfssone
  \newmathalphabet{\mathit}
  \addtoversion{normal}{\mathit}{cmr}{m}{it}
  \addtoversion{bold}{\mathit}{cmr}{bx}{it}
  \newmathalphabet{\mathbfit} 
  \addtoversion{normal}{\mathbfit}{cmr}{bx}{it}
  \addtoversion{bold}{\mathbfit}{cmr}{bx}{it}
  \newmathalphabet{\mathbfss} 
  \addtoversion{normal}{\mathbfss}{cmss}{bx}{n}
  \addtoversion{bold}{\mathbfss}{cmss}{bx}{n}
  \ifAMStwofonts
    \ifCUPmtlplainloaded \else
      \UseAMStwoboldmath
      \makeatletter
      \new@mathgroup\upmath@group
      \define@mathgroup\mv@normal\upmath@group{eur}{m}{n}
      \define@mathgroup\mv@bold\upmath@group{eur}{b}{n}
      \edef\UPM{\hexnumber\upmath@group}
      \new@mathgroup\amsa@group
      \define@mathgroup\mv@normal\amsa@group{msa}{m}{n}
      \define@mathgroup\mv@bold\amsa@group{msa}{m}{n}
      \edef\AMSa{\hexnumber\amsa@group}
      \makeatother
      \mathchardef\upi="0\UPM19
      \mathchardef\umu="0\UPM16
      \mathchardef\upartial="0\UPM40
      \mathchardef\leqslant="3\AMSa36
      \mathchardef\geqslant="3\AMSa3E

      \let\leq=\leqslant 
      \let\geq=\geqslant 
    \fi
  \fi
\fi 

\ifnfsstwo
  \DeclareMathAlphabet{\mathbfit}{OT1}{cmr}{bx}{it}
  \SetMathAlphabet\mathbfit{bold}{OT1}{cmr}{bx}{it}
  \DeclareMathAlphabet{\mathbfss}{OT1}{cmss}{bx}{n}
  \SetMathAlphabet\mathbfss{bold}{OT1}{cmss}{bx}{n}
  \ifAMStwofonts
    \ifCUPmtlplainloaded \else      \DeclareSymbolFont{UPM}{U}{eur}{m}{n}
      \SetSymbolFont{UPM}{bold}{U}{eur}{b}{n}
      \DeclareSymbolFont{AMSa}{U}{msa}{m}{n}
      \DeclareMathSymbol{\upi}{0}{UPM}{"19}
      \DeclareMathSymbol{\umu}{0}{UPM}{"16}
      \DeclareMathSymbol{\upartial}{0}{UPM}{"40}
      \DeclareMathSymbol{\leqslant}{3}{AMSa}{"36}
      \DeclareMathSymbol{\geqslant}{3}{AMSa}{"3E}

      \let\leq=\leqslant 
      \let\geq=\geqslant 
    \fi
  \fi
\fi 

\ifCUPmtlplainloaded \else
  \ifAMStwofonts \else 
    \def\upi{\pi}
    \def\umu{\mu}
    \def\upartial{\partial}
  \fi
\fi

\title{Star clusterings in the Carina complex:\\
$UBVRI$ photometry of Bochum~9, 10 and 
11}
\author[Ferdinando Patat \& Giovanni Carraro]
       {Ferdinando Patat$^{1\thanks{e-mail~:~fpatat@eso.org}}$ and  Giovanni Carraro$^{1,2}$\\         
$^{1}$ European Southern Observatory, Karl-Schwarzschild-Str 2, D-85748
Garching b. M\"unchen, Germany\\  
$^{2}$ Dipartimento di Astronomia, Universit\'a di Padova, Vicolo
Osservatorio
3 , I-35122 Padova, Italy}

\date{Accepted .
      Received;
      in original form }

\pagerange{\pageref{firstpage}--\pageref{lastpage}}
\pubyear{2001}

\begin{document}

\maketitle

\label{firstpage}

\begin{abstract}
We report on the first $UBVRI$ CCD photometry of three poorly known
star clusterings in the region of $\eta$~Carinae: Bochum~9,
Bochum~10 and Bochum~11. 
We found that they  are young, rather poor and loose
open clusters.\\
We  argue that Bochum~9 is probably a small and loose open
cluster with about 30 probable members
having E$(B-V)$~=~0.63$\pm$0.08, located 4.6 kpc far from the Sun,
beyond the Carina spiral arm.\\
Similarly, Bochum~10 is a sparse 
aggregate with 14 probable members having E$(B-V)$~=~0.47$\pm$0.05
and at a distance of 2.7 kpc from the Sun.\\ 
Finally, Bochum~11 is a less than  $4 \times 10^{6}$ yrs 
old cluster for which we identify 24 members. 
It has a reddening E$(B-V)$~=~0.58$\pm$0.05,
and lies between Bochum~10 and 9, at 3.5 kpc from the Sun. 
We propose that in the field of the cluster
some stars  might be pre Main Sequence (MS) candidates.

\end{abstract}

\begin{keywords}
Open clusters and associations--Individual: Bochum~9; Open clusters and
associations--Individual: Bochum~10; Open clusters and
associations--Individual: Bochum~11;
\end{keywords}

\section{Introduction}
Feinstein (1995) suggests that there are
14 known or suspected open star clusters probably
related to the Carina spiral feature.
Most of them are actually very well known.\\
Nonetheless some remain very poorly studied, and 
our knowledge often does not extend beyond the simple
identification. As a consequence
the real nature of some of these star
clusterings and
their belonging to the Carinae complex
are  not well established.\\
The region of $\eta$~Carinae on the other hand is since
long time recognized as an ideal laboratory to study the formation
of star clusters by obtaining precise photometry and age
estimates (Massey and Johnson 1993).\\
Having this in mind, we have undertaken a photometric survey
aimed at obtaining homogeneous high quality CCD data for all
the known star clusterings in this region.\\
We already reported about NGC~3324 and Loden~165
(Carraro et al 2001), showing that Loden~165 does 
probably not
belong to the Carina complex.\\
Here we report on three poorly studied objects:
Bochum~9, Bochum~10 and Bochum~11, 
identified in the seventies by Moffat \& Vogt (1975). Only
photoelectric photometry is available for these
objects;
in particular Bochum~9 deserves special attention
since it is not clear whether it is a cluster or not.
The other two clusters are young objects, with some
evidence of a pre MS population (Fitzgerald \& Mehta 1987).
Their basic parameters are given in Table~1.\\ 

\noindent
The plan of the paper is as follows. In Section~2 we
describe the data acquisition and reduction, while 
Sections 3 to 5 are dedicated to Bochum~9, Bochum~10
and Bochum~11, respectively. Finally Section~6
summarizes our results, and in the Appendix we provide
some additional details on the data reduction and
photometric errors.

\begin{table}
\caption{Basic parameters of the observed objects.
Coordinates are for J2000.0 equinox.}
\begin{tabular}{lcccc}
\hline
\hline
\multicolumn{1}{l}{Name} &
\multicolumn{1}{c}{$\alpha$}  &
\multicolumn{1}{c}{$\delta$}  &
\multicolumn{1}{c}{$l$} &
\multicolumn{1}{c}{$b$} \\
\hline
& $hh:mm:ss$ & $^{o}$~:~$^{\prime}$~:~$^{\prime\prime}$ & $^{o}$ & $^{o}$ \\
\hline
Bochum~9      & 10:35:45.7 & -60:07:33.8 & 286.80 & -1.58\\
Bochum~10     & 10:42:14.2 & -59:08:43.7 & 287.02 & -0.33\\
Bochum~11     & 10:47:15.2 & -60:05:50.8 & 288.04 & -0.87\\
\hline\hline
\end{tabular}
\end{table}

\begin{figure}
\centerline{\psfig{file=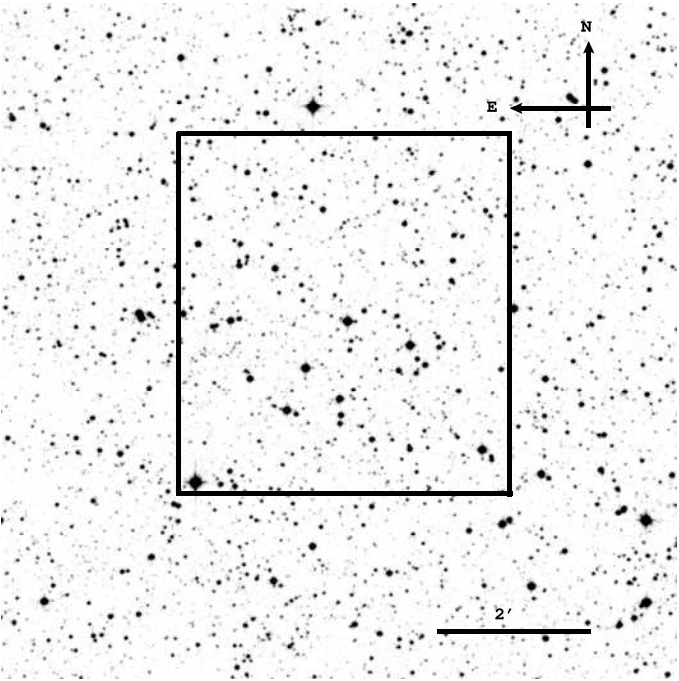,width=9cm,height=9cm}}
\caption{DSS map of a region around Bochum~9. The box confines the field
covered by our photometry.}
\end{figure}

\begin{table*}
\tabcolsep 0.10truecm
\caption{Journal of observations of Bochum~9 (April 13, 1996),
Bochum~10 (April 14 , 1996), and Bochum~11 (April 15, 1996).}
\begin{tabular}{cccccccccccccccc} \hline
      & \multicolumn{7}{c}{Bochum~9}         & \multicolumn{4}{c}{Bochum~10}         & \multicolumn{4}{c}{Bochum~11} \\
Field & Filter & Exp. Time & Seeing &  Field & Filter & Exp. Time &
Seeing  & Field        & Filter & Exp. Time & seeing & Field          & Filter & Exp. Time & seeing\\
      &        & (sec)     & ($\prime\prime$)&  &        & (sec)     & ($\prime\prime$)&  &     &  (sec)    & ($\prime\prime$) &        &  &(sec)    & ($\prime\prime$)\\
      &        &           &                 &        &           &                  &        &           & \\  
 $\#1$   & U &   30 &  1.7 &  $\#3$  & U &   30 &  1.9 & $\#1$ &   &      &      & $\#1$ &     &   &         \\
	 & U &  600 &  1.7 &         & U &  600 &  1.9 &       & U &   60 &  2.1 &      &U &  30 & 1.8\\
	 & B &   10 &  1.7 &         & B &   10 &  1.8 &       & B &   10 &  2.0 & &B &   5 & 1.8\\
	 & B &  600 &  1.3 &         & B &  600 &  1.9 &       & B &  300 &  1.9 & &B & 300 & 1.7\\
	 & V &    5 &  1.3 &         & V &    5 &  1.7 &       & V &    3 &  1.6 & &V &   3 & 1.6\\
	 & V &  300 &  1.4 &         & V &  300 &  1.7 &       & V &  120 &  1.6 & &V & 120 & 1.7\\
	 & R &    8 &  1.6 &         & R &    5 &  1.7 &       & R &    3 &  1.7 & &R &   3 & 1.7\\
	 & R &  180 &  1.6 &         & R &  180 &  1.7 &       & R &   60 &  1.5 & &R &  60 & 1.6\\
	 & I &   10 &  1.4 &         & I &    5 &  1.6 & & I &    5 &  1.5 & &I &   3 & 1.5\\
	 & I &  300 &  1.5 &         & I &  300 &  1.7 & & I &  120 &  1.5 & &I & 120 & 1.5\\
 $\#2$   & U &   30 &  1.7 &  $\#4$  & U &   30 &  2.0 & $\#2$ &   &      & &$\#2$ &    &   &         \\
	 & U &  600 &  1.7 &         & U &  600 &  1.7 & & U &   60 &  1.5 & &U &  30 & 2.0\\
	 & B &   10 &  1.8 &         & B &   10 &  1.5 & & B &   10 &  1.5 & &B & 300 & 2.0\\
	 & B &  600 &  2.1 &         & B &  600 &  1.6 & & B &  300 &  1.9 & &B &   5 & 1.9\\
	 & V &    5 &  1.8 &         & V &    5 &  1.8 & & V &    3 &  1.6 & &V & 120 & 1.9\\
	 & V &  600 &  1.7 &         & V &  300 &  1.7 & & V &  120 &  2.3 & &V &   3 & 1.8\\
	 & R &    8 &  1.7 &         & R &    5 &  1.7 & & R &    3 &  2.0 & &R &  60 & 1.6\\
	 & R &  180 &  1.7 &         & R &  180 &  1.8 & & R &   60 &  1.9 & &R &   3 & 1.6\\
	 & I &   10 &  1.7 &         & I &    5 &  1.9 & & I &    5 &  1.9 & &I & 120 & 1.6\\
	 & I &  300 &  1.7 &         & I &  300 &  1.9 & & I &  120 &  1.7 & &I &   3 & 1.6\\ 
\hline
\end{tabular}
\end{table*}

\section{Data acquisition and reduction}
Observations were conducted at La Silla on April 13-16, 1996, using the 
imaging Camera (equipped with a TK coated 512 $\times$ 512 pixels CCD \#33) 
mounted at the Cassegrain focus of the 0.92m ESO--Dutch telescope. All 
nights were 
photometric with a seeing ranging from 1$^{\prime\prime}$ to 
2$^{\prime\prime}$. 
The scale on the chip is 0$^{\prime\prime}.44$ and the array covers
about 3$^\prime$.3 $\times$ 3$^\prime$.3 on the sky. Due to the projected
diameter of the objects and the relatively small field of view,  
it was necessary to observe four fields for the same object.
To allow for a proper photometric calibration and to asses the night quality, 
the standard fields RU149, PG1323, PG1657, SA~109 and SA~110 (Landolt 1992)
were monitored each night. Finally, a series of flat--field frames on the
twilight sky were taken. The scientific exposures have been flat--field and
bias corrected by means of standard routines within {\it IRAF}\footnote{IRAF is
distributed by the National Optical Astronomy Observatories, which is
operated by the Association of Universities for Research in Astronomy,
Inc., under contract to the National Science Foundation.}. Further 
reductions were performed using the DAOPHOT package (Stetson 1991) in the
{\it IRAF} environment.

The instrumental magnitudes have been transformed into standard Bessel $UBVRI$
magnitudes using fitting coefficients derived from observations of the
standard field stars from Landolt (1992), after including exposure time
normalization and airmass correction. Aperture corrections have
also been applied. The observations log-book  for the three
clusters is presented in Table~2, whereas additional details of the
photometric reduction and error analysis are provided in the Appendix.

\begin{table*}
\caption{Photometry and astrometry of the brightest stars in the field
of Bochum~9. Absolute proper motion are taken from Tycho~2, and are expressed
in $mas/yr$.}
\begin{tabular}{cccccccccc}
\hline
\hline
\multicolumn{1}{c}{ID} &
\multicolumn{1}{c}{Star}  &
\multicolumn{1}{c}{$V$}  &
\multicolumn{1}{c}{$(B-V)$} &
\multicolumn{1}{c}{$(U-B)$} &
\multicolumn{1}{c}{$\mu_{\alpha*}$}  &
\multicolumn{1}{c}{$\sigma_{\mu}$}  &
\multicolumn{1}{c}{$\mu_{\delta}$} &
\multicolumn{1}{c}{$\sigma_{\mu}$} &
\multicolumn{1}{c}{$Sp. Type$} \\
\hline
1  &                 &  9.381 & -0.003 & -0.441 &        &      &      &      &    \\
2  &  HD~305364      &  9.767 &  0.025 & -0.028 & -12.90 & 5.30 & 1.10 & 1.90 & B8 \\
3  &  HD~305368      & 10.188 &  0.332 & -0.645 & -24.00 &      & 1.00 &      & A0 \\
4  &  HD~305366      & 10.828 &  0.017 & -0.132 & -16.00 &      & 0.00 &      & G8 \\
5  &  GSC08957-02702 & 11.071 &  0.123 &  0.205 & -14.30 & 5.30 & 6.50 & 2.40 & ?  \\
6  &                 & 11.388 &  0.339 & -0.577 &        &      &      &      &    \\
7  &  HD~305383      & 11.394 &  0.367 & -0.633 & -12.50 & 7.50 & 2.50 & 1.80 & A  \\
8  &                 & 11.531 &  0.316 &  0.326 &        &      &      &      &    \\
10 &  HD~305370      & 11.971 &  0.227 &  0.310 & -25.00 &      & 0.00 &      & B9 \\ 
   &  HD~91025       &        &        &        &  -8.10 & 5.30 & 3.20 & 3.60 & B1 \\
\hline\hline
\end{tabular}
\end{table*}

\section{ Bochum~9}\label{sec:bo9}
Bochum~9 appears as a sparse group of 7 blue luminous stars
westwards of $\eta$~Carinae. This star clustering was identified
and studied by Moffat \& Vogt (1975) who provided photo-electric
photometry for 22 stars. They concluded that 12 of them are O and B
type stars, but the sample seems not to show any sequence, and this
raises some doubts about the actual nature of this group. 
To our knowledge  no other studies have been carried out so far.\\

\begin{figure}
\centerline{\psfig{file=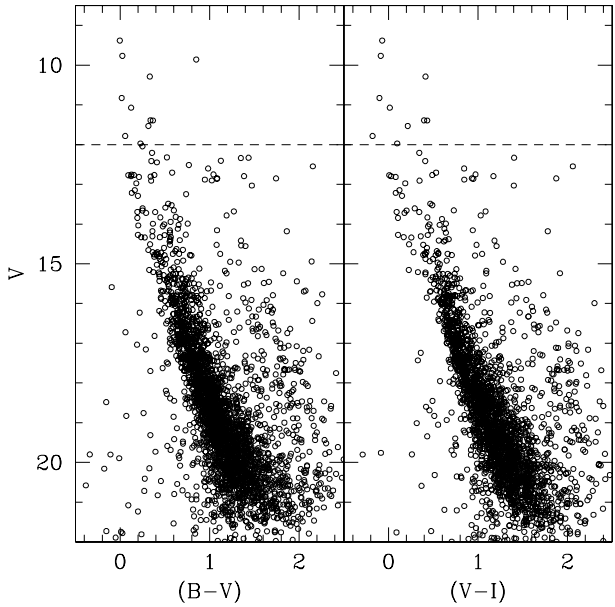,width=9cm,height=9cm}}
\caption{CMDs for all stars in the region of Bochum~9.
The dashed line indicates the limiting magnitude reached by previous
investigations.}
\end{figure}

\begin{figure}
\centerline{\psfig{file=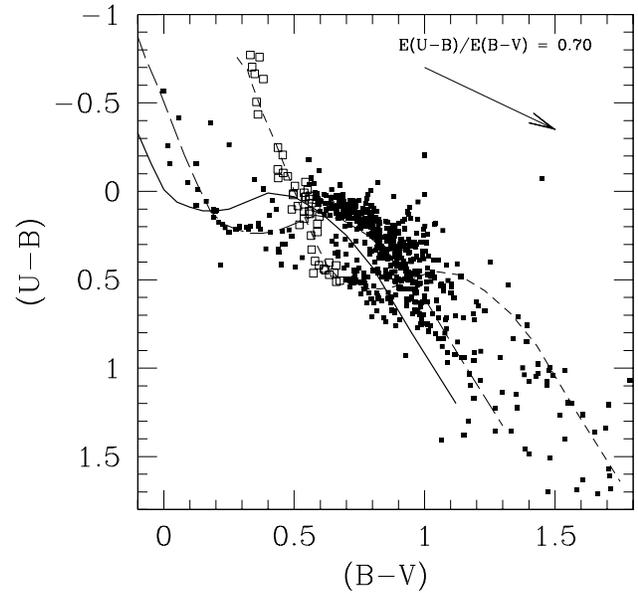,width=9cm,height=9cm}}
\caption{Two colors diagram  for all stars in the region of
Bochum~9 having $V \leq$ 16.5. 
The arrow indicates the reddening vector. The solid
line is the empirical Schmidt-Kaler (1982) ZAMS. Open squares
indicate stars having E$(B-V)$=0.63$\pm$0.08, and the dashed line
crossing these stars is an empirical ZAMS shifted by  E$(B-V)$=0.63.
Most of the other stars lie along an empirical ZAMS (long dashed line)
shifted by  E$(B-V)$=0.18. See tex for additional details.}
\end{figure}

We obtained $UBVRI$ CCD photometry for about 4000 stars in the region
of Bochum~9 down to $V$~=~22. 
We have 9 stars in common with Moffat \& Vogt (1975),
which are listed in Table~3. The mean difference turns out to be:

\[
V_{PC} - V_{MV} = 0.006\pm0.018 
\]

\[
(B-V)_{PC} - (B-V)_{MV} = 0.024\pm0.015 
\]

\[
(U-B)_{PC} -(U-B)_{MV} = 0.030\pm 0.025
\]

\noindent
The agreement is very good both for magnitude and for colors.\\

The CMDs for all stars detected in the region of Bochum~9 are
shown in Fig.~2. The most prominent feature is a well defined MS from
$V$~=~10 up to $V$~=~22. The stars on the red side of the MS appear
to form a sequence parallel to the MS, that can be ascribed to the He-burning
evolved stars (Red Giant Branch (RGB) stars) projected
onto  the line of sight with increasing reddening
(see the {\it F3} field CMD in Vallenari et al 2000, which lies very close to Bochum~9
and has a pretty similar appearance).
Already Moffat \& Vogt (1975) from their small sample,  suggested
that probably we are considering stars located at different distances
in the direction of Carina spiral arm, in a region of small
reddening.

\begin{figure}
\centerline{\psfig{file=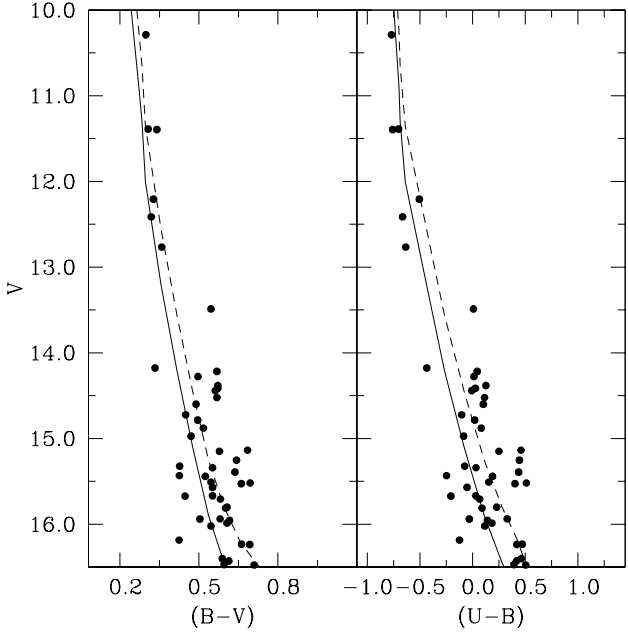,width=9cm,height=9cm}}
\caption{CMDs for all the
stars in the region of Bochum~9 having $V \leq$ 16.5
and sharing the reddening E$(B-V)$~=~0.63$\pm$0.08. 
The empirical
ZAMS from Schmidt-Kaler (1982) is shown as a solid line
and as been shifted according to the reddening and distance modulus.
The dashed line represents the same ZAMS 0.75 mag brighter, which
defines the locus of unresolved binaries.
See the text for more details}
\end{figure}

\begin{figure}
\centerline{\psfig{file=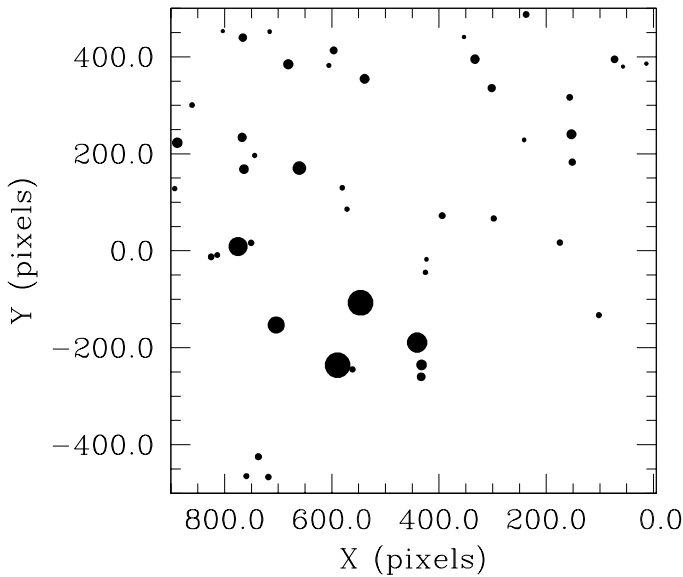,width=9cm,height=9cm}}
\caption{The projected spatial distribution of the candidate cluster
members in the field of Bochum 9. The size of the circles
is proportional to the magnitude of the stars. The field is about  
$6^{\prime} \times 6^{\prime}$.}
\end{figure}

\begin{figure}
\centerline{\psfig{file=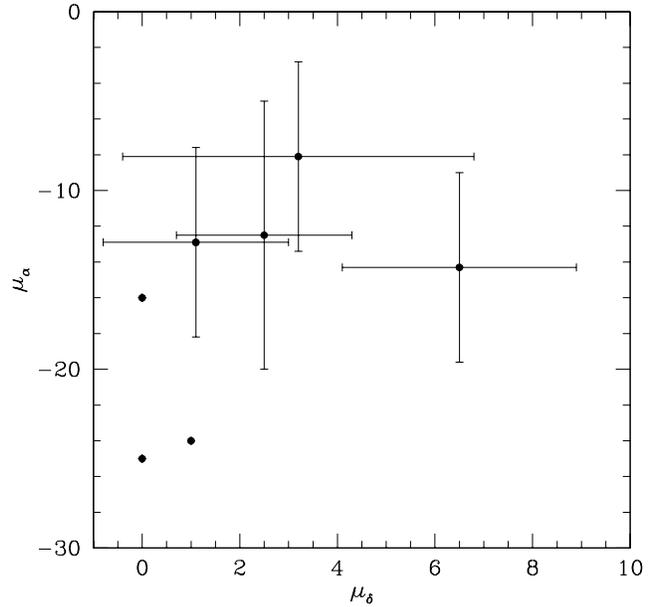,width=9cm,height=9cm}}
\caption{Vector point diagram for Tycho~2 stars in the
field of Bochum~9. Proper motions are expressed in $mas/yr$,
and crosses, where available, indicate the uncertainties.}
\end{figure}

\begin{table}
\tabcolsep 0.15cm
\caption{Photometry of the brightest stars in the field
of  the open cluster Bochum~10.}
\begin{tabular}{ccccccc}
\hline
\hline
\multicolumn{1}{c}{ID} &
\multicolumn{1}{c}{$V$}  &
\multicolumn{1}{c}{$(B-V)$}  &
\multicolumn{1}{c}{$(U-B)$} &
\multicolumn{1}{c}{$(V-R)$} &
\multicolumn{1}{c}{$(V-I)$}  \\
\hline
 1  &       9.349&    0.187&   -0.744&    0.020&    0.135\\
 2  &       9.292&    0.063&    0.029&   -0.064&   -0.068\\
 3  &       9.503&    0.076&   -0.801&   -0.026&    0.015\\
 4  &       9.553&    0.106&   -0.758&   -0.028&    0.030\\
 5  &       9.603&    0.066&   -0.702&   -0.069&   -0.066\\
 6  &       8.782&    1.482&    1.650&    0.588&    9.999\\
 7  &      10.194&    0.099&   -0.725&   -0.032&    0.018\\
 8  &      10.550&    0.160&   -0.677&   -0.001&    0.096\\
 9  &      11.172&    1.142&    0.973&    0.456&    0.976\\
 10 &      11.563&    0.167&    0.160&   -0.067&   -0.007\\
 11 &      11.871&    0.668&    0.088&    0.292&    0.605\\
 12 &      11.959&    0.541&    0.022&    0.209&    0.459\\ 
 13 &      12.265&    0.233&   -0.382&    0.050&    0.179\\
 14 &      12.177&    0.436&    0.122&    0.140&    0.333\\
 15 &      12.227&    0.847&    0.338&    0.359&    0.728\\
 16 &      12.598&    0.411&   -0.389&    0.205&    0.491\\
\hline
\end{tabular}
\end{table}

\subsection{Color-color and color-magnitude diagrams}
The position of all the stars brighter than $V$~=~16.5 in the color-color
diagram is shown in  Fig.~3.
The solid line is the un-reddened Schmidt-Kaler (1982)
sequence.\\
There seems to be two distinct populations\\
The bulk of the stars are indicated with filled squared, and 
are fitted by an empirical ZAMS (long dashed line)
shifted by E$(B-V)$~=~0.18.
This sequence is probably due to stars
in the Galactic disk located between us and the Carina spiral arm.\\
There is however another group of stars, plotted with open squares,
sharing a common reddening E$(B-V)$~=~0.63$\pm$0.08.
They are 48 stars.\\
The sequence (dashed line) crossing these stars is the same empirical ZAMS
as above, but shifted by E$(B-V)$~=~0.63.\\

We shall refer to the more reddened sequence as Bochum~9 
candidate members.
In order to better understand the nature of Bochum~9 we plotted
the cluster candidate members in the color-magnitude diagrams
shown in Fig.~4. The selected stars define a tight sequence
in the $V$ vs $(B-V)$ plane (Fig.~4, left panel), where we super-imposed
an empirical ZAMS shifted by E$(B-V)$~=~0.63 and 
by the apparent distance modulus $(m-M)$~=~15.3 .
In the right panel we present the $V$ vs $(U_B)$
CMD for the same group of stars. It seems that they still follow  
a sequence, although somewhat scattered below $V$~=~15.3.
In this panel the ZAMS has been shifted by E$(U-B)$~=~0.44 
and by the apparent distance modulus $(m-M)$~=~15.3.
The appearance of these CMDs suggests that of Bochum~9 is
a probable very young loose open cluster located about 4.6 kpc
far from the Sun, beyond the Carina spiral arm.
To better identify cluster members we assumed that some of the
stars in Fig.~4 might be  binaries. Unresolved binaries populate a sequence
0.75 mag brighter than the single stars ZAMS, offering us the
possibility to define the width of the MS. The binaries sequence
has been drawn in Fig~4 as a dashed line. 
As a consequence, if we consider as non member all the stars
lying out of the region defined by the two ZAMS, the number of candidate
members turns out to be 29. \\

The projected spatial distribution of the candidate members
in shown in Fig.~5. The stars do not show any cluster like
distribution, raising some doubts on the cluster nature of Bochum~9.

\subsection{Astrometric data}
For 7 of the most luminous stars we have at disposal the absolute
proper motions provided by Tycho~2 (H\o g E.\ et al.\, 2000), and they are
listed in Table~3. First of all. We want to stress that the  
uncertainties in the measurements (due to 
the fairly large distance) cannot help
too much , to draw any firm conclusions. Anyhow,
while they appear broadly consistent in
$\mu_{\delta}$ (except for GSC08957-0270), they seem to form
two sub-groups in $\mu_{\alpha*}$. HD~305364 shares the same  $\mu_{\alpha*}$
of HD~305383, HD~92025 and maybe HD~305366, while 
HD~305368 has basically the same $\mu_{\alpha*}$ of HD~305770.
The star GSC08957-02702 probably is an isolated case.\\

\noindent
The available data do not allow
us to decide unambiguously about the nature of Bochum~9.
It is clear that the Galactic disk component is rather strong in this 
direction (Vallenari et al 2000).
Anyhow, we believe that we are probably facing
a young loose open cluster about 4.6 kpc far from the Sun,
at odds with previous suggestions by Moffat \& Vogt (1975).\\
We emphasize that proper motions
provide  only a weak suggestion, and that  radial
velocities for the brightest stars or a dedicated relative proper
motions survey are needed to decide
about the real nature of Bochum~9.\\

\noindent
The position in the CMDs (see Fig.~4) of some stars
below $V$~=~14.5 deserves a final consideration.
The most plausible explanation for the scatter in this region
is that these stars can be field stars. However,
There is the possibility that some of them might be pre MS
stars, an hypothesis  which
demands further investigations by $H_{\alpha}$ 
equivalent width measurements 
or infrared photometry (see for instance
Vallenari et al 1999 and the discussion for Bochum~11 in Sec.~5.2).

\begin{table*}
\caption{Photometric and astrometric data for the brightest 
stars in the region of Bochum~10. 
Data is taken from Tycho 2. Fe indicates Feinstein (1981)
numbering, while $m$ stands for members, as identified
by Feinstein.}
\begin{tabular}{l@{\hspace{2pt}}lllcccccccc}
\hline
\hline
\multicolumn{1}{c}{ID} &
\multicolumn{1}{c}{Fe} &
\multicolumn{1}{c}{Star} &
\multicolumn{1}{c}{$\mu_{\alpha*}$}  &
\multicolumn{1}{c}{$\sigma_{\mu}$}  &
\multicolumn{1}{c}{$\mu_{\delta}$} &
\multicolumn{1}{c}{$\sigma_{\mu}$} &
\multicolumn{1}{c}{$V$} & 
\multicolumn{1}{c}{$(B-V)$} & 
\multicolumn{1}{c}{$(U-B)$} & 
\multicolumn{1}{c}{$Sp. Type$} \\
\hline
1 &  5m & HD~92894  &  -2.30  & 3.60 &  1.40 & 1.60 &  9.394 & 0.187 & -0.744 & B01V  \\
2 &  6  & HD~92909  & -15.90  & 2.10 & -0.20 & 1.90 &  9.292 & 0.063 &  0.029 & B9.5V \\
3 &  4m & HD~303297 &   0.40  & 4.40 & -1.20 & 3.20 &  9.503 & 0.076 & -0.801 & A2    \\
4 &  3m & HD~303296 &  -0.10  & 3.60 &  2.80 & 2.70 &  9.553 & 0.106 & -0.758 & Be    \\
5 & 15m & HD~92852  & -13.50  & 8.70 &  0.60 & 1.80 &  9.603 & 0.066 & -0.702 & B4    \\
6 & 14  & HD~92835  & -15.50  & 9.60 & -1.50 & 2.60 &  8.782 & 1.482 &  1.650 & K2III \\
7 &  8m & HD~302989 & -11.00  &      & -4.00 &      & 10.194 & 0.099 & -0.725 & A    \\
  &  9  & HD~303188 &  -6.10  & 6.70 & -0.60 & 4.30 & & & & B3    \\
  & 10m & HD~92739  & -12.60  & 2.90 &  2.50 & 2.60 & & & & B1/B2 \\
  & 11m & HD~303190 & -12.70  & 7.40 &  2.50 & 3.10 & & & & B5    \\
  & 23  & HD~303295 &  -9.00  &      & -3.00 &      & & & & B8    \\
  &  1m & HD~93002  &   1.10  & 2.90 &  5.60 & 2.10 & & & & B2III \\
  & 35m & HD~93026  &  -1.70  & 4.70 &  1.80 & 3.90 & & & & B2III \\
  & 18  & HD~303291 & -19.00  &      &  0.00 &      & & & & ?     \\
  & 33  & HD~93055  &  -6.00  & 2.90 &  5.30 & 1.60 & & & & B8/B9 \\
  & 34  & HD~93114  & -25.70  & 3.00 &  5.50 & 2.30 & & & & A7II  \\
  & 38  & HD~303290 & -13.00  &      & -2.00 &      & & & &  ?    \\
\hline\hline
\end{tabular}
\end{table*}

\begin{figure}
\centerline{\psfig{file=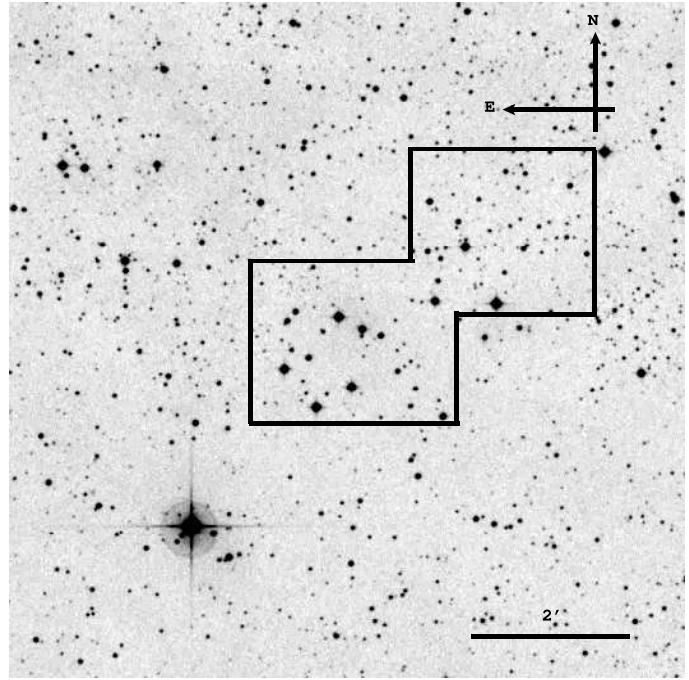,width=9cm,height=9cm}}
\caption{DSS map of a region around Bochum~10. The solid line confines the field
covered by our photometry.}
\end{figure}

\section{ Bochum~10}\label{sec:bo10}
Bochum~10 is a loose  open cluster located  north-west
of $\eta$~Carinae, and surrounded by a diffuse nebulosity
(Smith et al 2000). We therefore expect to find
signatures of variable extinction across
the cluster.
Bochum~10 was firstly reported by Moffat \& Vogt (1975) who obtained $UBV$
photoelectric photometry for 17 stars, and found 12 possible
members. 
Later Feinstein (1981) obtained photoelectric $UBVRI$
photometry for 39 stars down to $V$~=~13, recognizing 15 members. 
Finally, Bochum~10 has been studied by Fitzgerald \& Mehta (1987)
who provide photoelectric $UBV$ photometry of about twenty
stars down to V=13.5.
They suggest that the cluster is rather sparse and includes O and
B stars only. No members are suggested to exist below $V$~=~13.5.\\
While all these studies give the same estimate of the reddening,
being E$(B-V) \approx$  0.35, there are some discrepancies in the distance
modulus, which ranges from $(m-M)_o$~=~12.03 to 12.80.
According to Moffat \& Vogt (1975) the cluster is 2.56 kpc far from
the Sun, whereas for Feinstein (1981) is 3.6 kpc distant and
for Fitzgerald \& Mehta (1987) about 2.8 kpc.
The precise distance of Bochum~10 is crucial to assess its
membership to the Carina complex.
As for the age, there is some agreement in the literature,
being the cluster as old as $7 \times 10^{6}$ yrs.\\

For this object we obtained $UBVRI$ photometry for about 600 stars down to 
$V$~=~20.
The comparison of our data with those of Feinstein (1981)
for 17 common  stars (see the details in Table~5),  yields the following results:

\[
V_{PC} - V_{F} = -0.05\pm0.05 
\]

\[
(B-V)_{PC} - (B-V)_{F} = 0.06\pm0.03 
\]

\[
(U-B)_{PC} -(U-B)_{F} = 0.05\pm 0.05
\]

\[
(V-R)_{PC} - (V-R)_{F} = -0.09\pm0.02 
\]

\[
(V-I)_{PC} -(V-I)_{F} = -0.12\pm 0.03
\]

\noindent
Our field (see Fig.~6) comprises the region defined by
the 7 brightest stars mentioned by Moffat \& Vogt(1975),
but it is much smaller than the fields studied by Feinstein (1981)
and Fitzgerald \& Mehta (1987). By inspecting a DSS map
it appears however difficult to define the cluster center, since
there are several  bright stars evenly distributed
within a radius of 20$^{\prime}$  from the assumed
cluster center which seem to form separated clumps.
Anyhow, if we are really sampling the cluster core, our
photometry supersedes all previous studies and we can obtain
improved estimates of the cluster fundamental parameters.

\begin{figure}
\centerline{\psfig{file=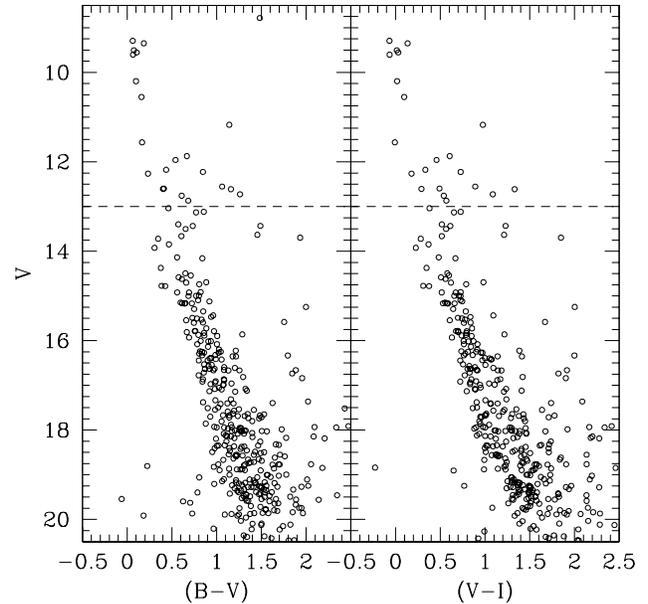,width=9cm,height=9cm}}
\caption{CMDs for all the stars in the region of Bochum~10.
The dashed line defines the limit of previous photometries.}
\end{figure}

\begin{figure}
\centerline{\psfig{file=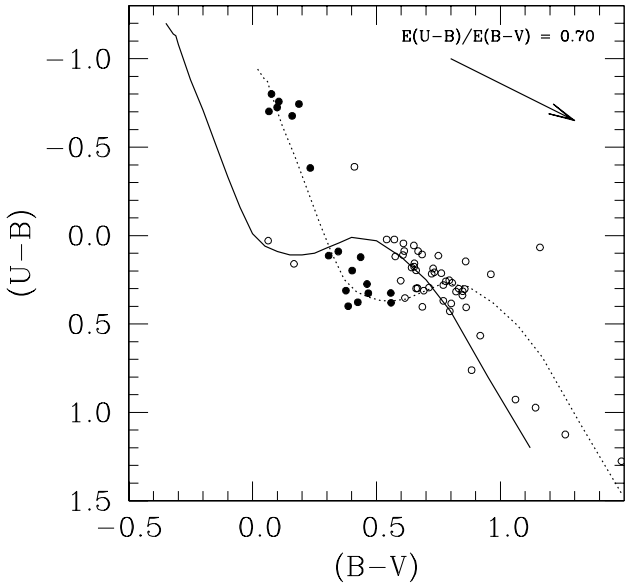,width=9cm,height=9cm}}
\caption{Two colors diagram  for all the stars in the region of
Bochum~10. The arrow indicates the reddening vector.
The solid line is the empirical Schmidt-Kaler (1982)
ZAMS, whereas the dotted line is the same ZAMS, but shifted by 
E$(B-V)$~=~0.47 . Filled circles indicate candidate members.}
\end{figure}

\begin{figure}
\centerline{\psfig{file=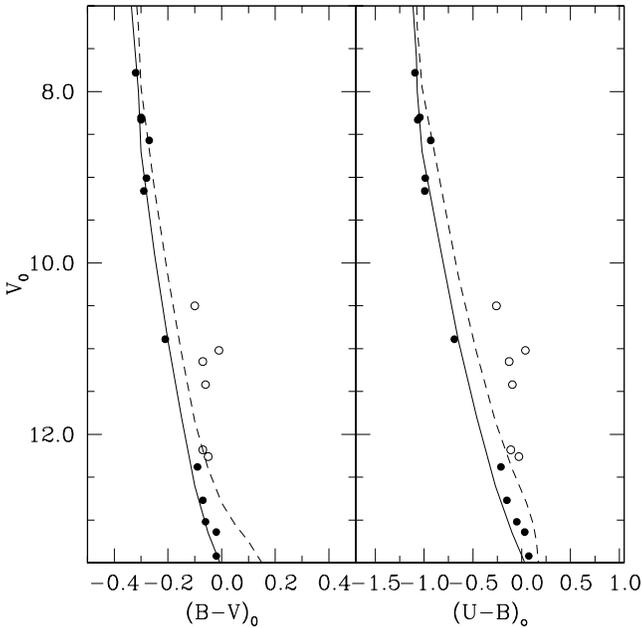,width=9cm,height=9cm}}
\caption{Reddening corrected CMDs for the 
candidate member stars in the region of Bochum~10. The solid line is the
empirical ZAMS from Schmidt-Kaler (1982), whereas
the dashed line is the same ZAMS, but shifted by 0.75 mag
to define the locus of unresolved binaries. 
Open circles indicate probable non members. See text for details}
\end{figure}

\subsection{Color-magnitude diagrams}
The CMDs for all detected stars in the field of Bochum~10
are shown in Fig.~8, where the dashed line indicates the limiting
magnitude of previous investigations.
The CMD looks very similar to that of Bochum~9, shown in Fig.~2.
Since the field is much smaller, there are less stars, 
but their distribution is practically identical, with
a thin MS extending from $V$~=~9 to $V$~=~20.\\
The contribution  of field stars is negligible up to
$V$~=~14.0. At fainter magnitudes, the MS is dominated by
the Galactic disk population.
As in the case of Bochum~9, there are some indications of 
Galactic disk RG stars, at least looking
at the scarcely populated vertical sequence which departs
from the MS at $V$ about 20.\\
The most prominent difference with Bochum~9 is the presence
of a clump of stars in the bright end of the MS.

\begin{figure}
\centerline{\psfig{file=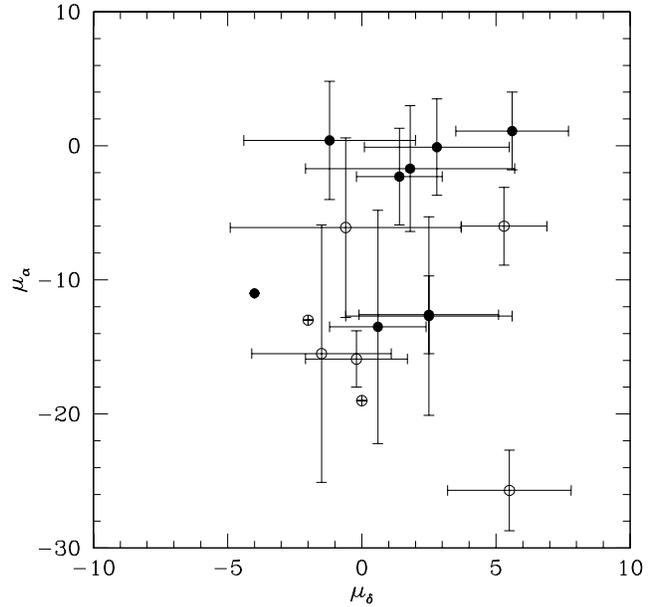,width=9cm,height=9cm}}
\caption{Vector point diagram for Tycho~2 stars in the
field of Bochum~10. Filled symbols indicate cluster members
according to Feinstein (1981), whereas open symbols indicate
cluster non-members. Crosses indicate the uncertainties
in the measurement. Proper motions are in $mas/yr$.}
\end{figure}

\subsection{Two color diagram}
All the stars having $UBV$ photometry have been
plotted in the two color digram in Fig.~9 .
Most of them exhibit basically no reddening, like the bulk of stars
in Bochum~9. However the clump of stars at $(B-V)$~=~0.1
and $(U-B)$~=~$-$0.7 has a larger reddening E$(B-V)$~=~0.47, and
is identified by an empirical ZAMS shifted by this amount
(dotted line). Since these stars have a common extinction, which is
significantly larger than that undergone by the bulk of the stars, 
it is reasonable to see whether they define a distinct sequence in 
the color magnitude diagram.\\
\noindent
For this purpose
we selected all the stars having 0.42 $\leq$ E$(B-V)$ $\leq$ 0.52
(18 stars in total).
These are plotted in Fig.~10, and
seem to actually define a clear sequence. There are 6  stars
which lie off the MS, and they are plotted with open circles.
The ZAMS has been over-imposed by adopting $(m-M)_o$~=~12.20,
which implies a distance of 2.7 kpc from the Sun.
The dashed line in Fig.~10 represents the same ZAMS 0.75 mag brighter,
and it defines the ZAMS of unresolved binaries.
Accordingly the two stars at $V_0 \approx 12.3$, $(B-V)_0 \approx -0.8$
may be binaries.

We looked for the relative position of these stars within the cluster, 
and found that, with the exception of the two binaries,
they lie in the outskirts of the area we covered.
Therefore we suggest that these stars do not probably belong to
Bochum~10.

\begin{table}
\tabcolsep 0.02cm
\caption{Photometry and astrometry of the brightest stars
if the field of Bochum~11. Proper motions are taken from Tycho
2 and are expressed in $mas/yr$.}
\begin{tabular}{cccccccccc}
\hline
\hline
\multicolumn{1}{c}{ID} &
\multicolumn{1}{c}{Name} & 
\multicolumn{1}{c}{$V$}  &
\multicolumn{1}{c}{$(B-V)$}  &
\multicolumn{1}{c}{$(U-B)$}  &
\multicolumn{1}{c}{$\mu_{\alpha*}$}  &
\multicolumn{1}{c}{$\sigma_{\mu}$}  &
\multicolumn{1}{c}{$\mu_{\delta}$} &
\multicolumn{1}{c}{$\sigma_{\mu}$} &
\multicolumn{1}{c}{$Type$} \\
\hline
1  & HD~93632   &  8.326 & 0.231 & -0.762 & -129.80 & 2.70 & 112.00 & 2.60 & O$^+$ \\
2  & HD~93576   &  9.661 & 0.163 & -0.727 &         &      &        &      &       \\
3  & HD~305612  & 10.288 & 0.270 & -0.729 &  -55.00 &      &  -3.00 &      & B     \\
4  & HD~93632B  & 10.489 & 0.122 & -0.711 &   -7.90 & 3.30 &   5.80 & 1.70 & B2V   \\
5  &            & 10.582 & 0.303 & -0.600 &         &      &        &      &       \\
6  & CPD~592704 & 10.841 & 0.347 & -0.596 &         &      &        &      &       \\
7  &            & 10.420 & 1.375 &  1.582 &         &      &        &      &       \\
8  &            & 11.321 & 0.080 & -0.702 &  -1.40  & 5.60 &   7.20 & 1.60 & B     \\ 
9  &            & 11.819 & 0.372 & -0.447 &         &      &        &      &       \\
\hline\hline
\end{tabular}
\end{table}

\noindent
\subsection{Astrometric data}
Tycho~2 provided measurements of proper motions for 17 stars
in the field of Bochum~10. They are listed in Table~6,
and plotted in the vector point diagram in Fig.~11.
Also in this case we stress that the uncertainties
in the proper motions are large. Nevertheless, the distribution
of the points in Fig.~11 is conducive to argue that 
two separate clumps with a large spread seem to exist, while
one object (HD~93114) clearly deviates from the bulk.\\
If these clumps are real, we are most likely looking at a group
of objects which do not share  the same motion properties. 
Interestingly Feinstein (1981) on a purely photometric basis attributed
membership to stars having discrepant motion (see Fig~11).
Finally, by closely inspecting the spatial distribution
of these stars in the sky (see Feinstein 1981, Fig.~1.)
it is possible to see how presumed members do not
distribute in the central region the cluster, but
members and non-members mix together.\\

\noindent
In conclusion, the results we obtained for Bochum~10
are the following. 
As in the case of Bochum~9, there are indications that 
we are not looking at a really bounded open cluster, but simply
at a loose group of bright stars embedded in a rich Galactic Disk field.\\
We found 14 probable members, less than Fitzgerald \& Metha (1987),
probably because  we have sampled a region too small to be really 
representative. The cluster seems to be quite young, with no stars in 
the act of leaving the MS.\\

\noindent
In order to better clarify the nature of Bochum~10, 
we suggest further investigations in two directions: a radial velocity survey 
for the brightest stars and a larger area for the photometric coverage.

\begin{figure}
\centerline{\psfig{file=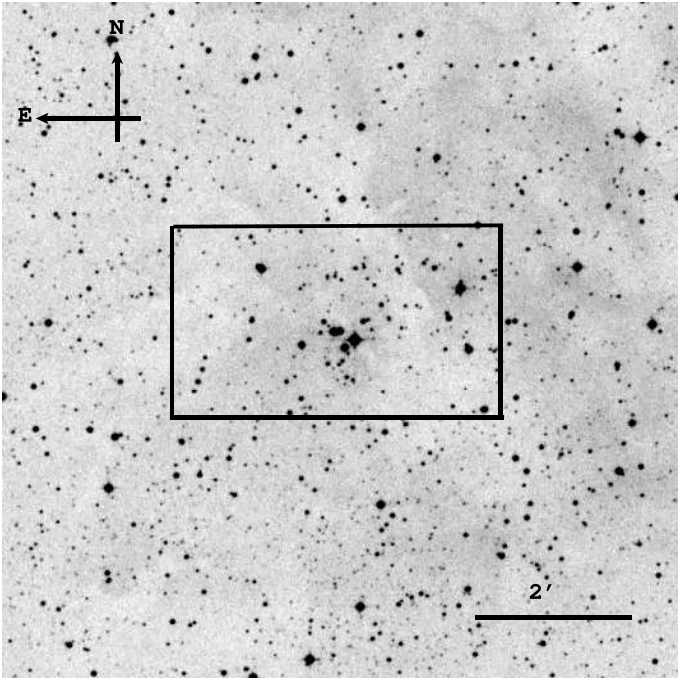,width=9cm,height=9cm}}
\caption{DSS map of a region around Bochum11. The box confines
the field covered by our observations.}
\end{figure}

\begin{figure}
\centerline{\psfig{file=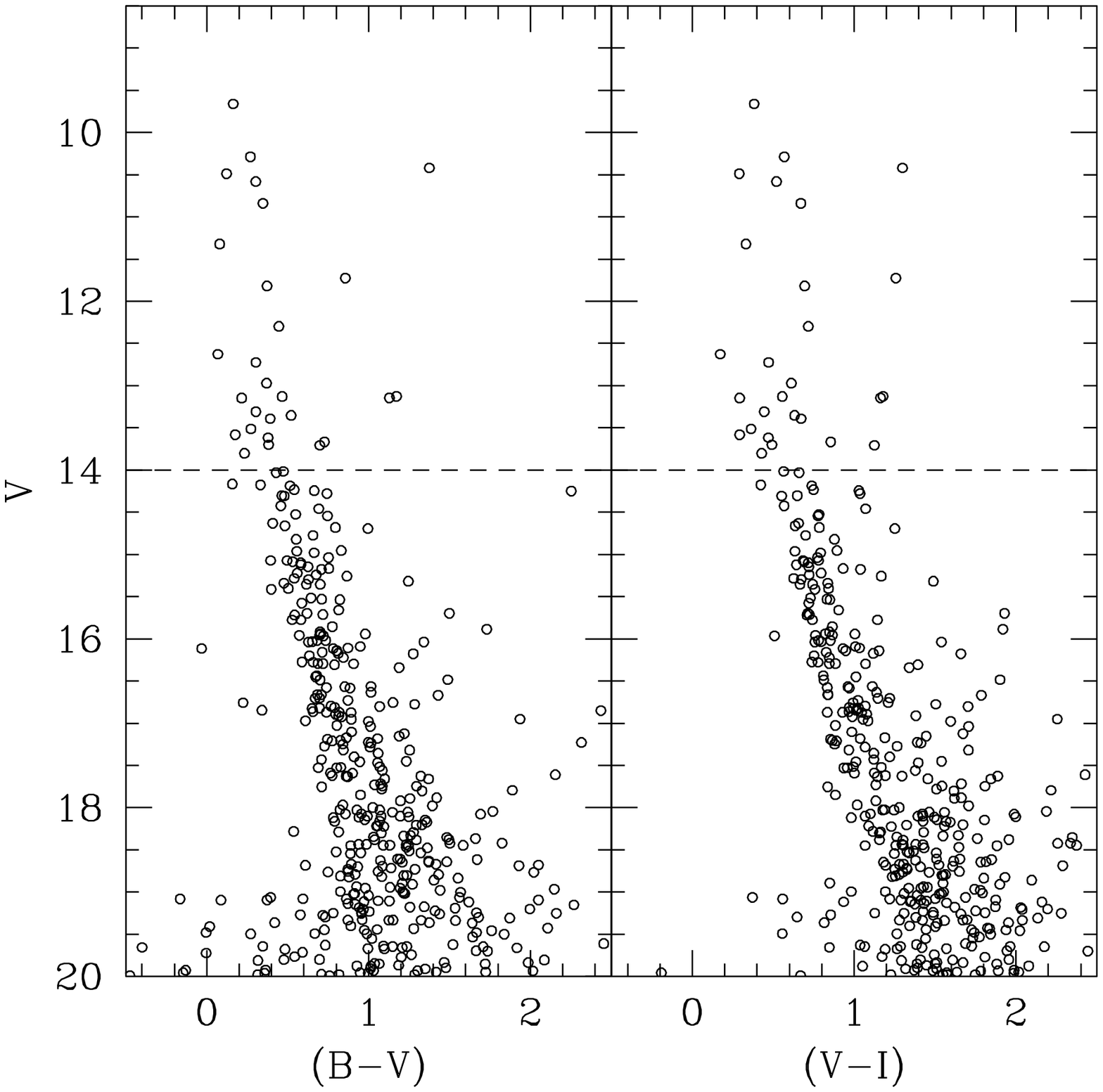,width=9cm,height=9cm}}
\caption{CMDs for all the stars in th region of Bochum~11.
The dashed line indicates the limiting magnitude of previous
investigations.}
\end{figure}

\section{ Bochum~11}\label{sec:bo11}
Bochum~11 is a compact very young group of stars 
around the double star HD~93962 (see Fig.~A1). 
The diffuse nebulosity (see Fig.~12) which surrounds Bochum~11 
is possibly what remains of  the cloud from which the cluster was
born. 
It was firstly investigated by Moffat
\& Vogt (1975), who provided $UBV$ photoelectric photometry for 8 probable
members. Soon after Forte (1976) obtained $UBV$ photoelectric photometry
for two stars already studied by Moffat \& Vogt (1975). 
Interestingly, Moffat \& Vogt (1975) pointed out that the cluster
suffers from differential reddening, being 
$E(B-V)~=~0.54\pm0.13$. Moreover they suggested that the stars fainter
than $V$~=~12 may be pre MS stars in contraction phase.\\
Finally, additional photoelectric  photometry has been carried out
by Fitzgerald \& Mehta (1987), who added 9 more stars.
By increasing to 15 the number of members, they 
confirmed that Bochum~11 is very young and pre MS stars
are possibly present below $V$~=~12.\\
A comparison with the work of  Fitzgerald \& Mehta (1987) for 9 stars
in common (see the details in Table~7),   yields the following results:

\[
V_{PC} - V_{FM} = -0.044\pm0.042 
\]

\[
(B-V)_{PC} - (B-V)_{FM} = -0.055\pm0.037 
\]

\[
(U-B)_{PC} -(U-B)_{FM} = -0.001\pm 0.029
\]

\noindent 
where PC refers to the present study, FM to Fitzgerald \& Mehta (1987).

\begin{figure*}
\centerline{\psfig{file=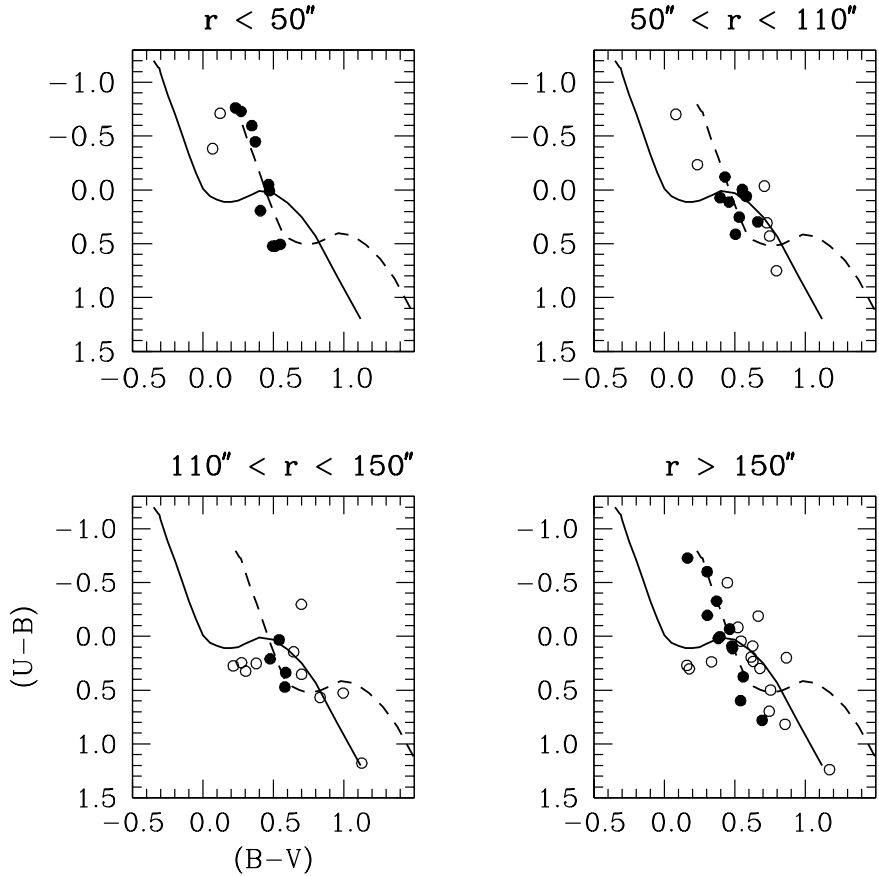,width=14cm,height=14cm}}
\caption{Two colors diagram  for the stars lying within different circles
centered in  Bochum~11. Solid circles identify candidate cluster
members.}
\end{figure*}

Our study covers a region of $3^{\prime} \times 6^{\prime}$ (see
Fig.~12) and we
obtained CCD $UBVRI$ photometry for about 780 stars down to $V$~=~21
(see Fig.~13), superseding all previous investigations. 
The CMD in Fig.~12 look quite similar to those of Bochum~9 and 10, 
so that the same kind of consideration hold also for Bochum~11 CMD.\\

\noindent
The data we collected
allow us to determine the cluster fundamental parameters.

\subsection{Cluster reddening}
We isolated Bochum~11 possible members on the basis of the position
in the two colors diagram  (see Fig.~14) and in the map in Fig.~A1.\\
By considering the stars distribution as projected onto the sky
(Fig.~A1), we selected four regions, 
namely a circle centered in HD~93632 with a radius of 50$^{\prime\prime}$
(upper left panel in Fig.~13), two  coronae defined by  
50$^{\prime\prime} \leq$ r $\leq 110^{\prime\prime}$  (upper righ panel)
and 110$^{\prime\prime} \leq$ r $\leq$150$^{\prime\prime}$ (lower left panel),
and finally the region outside the last corona, defined by 
r $\geq$150$^{\prime\prime}$.\\
In all these figures, the empirical Schmidt-Kaler (1982) ZAMS is plotted
as a solid line, whereas the same ZAMS, shifted by E$(B-V)$~=~0.58,
is drawn as a dotted line.
The central region seems indeed to be constituted by a group of 5-10
stars sharing the same reddening E$(B-V)$~=~0.58, whereas the majority
of stars lying outside the cluster core have basically no reddening,
like in the case of most of Bochum~9 and 10 stars.
It is reasonable to conclude that these are field stars.\\
We have indicated with filled circles candidate cluster members on
the basis of common reddening. Noticeably, we find 
probable cluster members also in the outer rings. In total,
the candidate members are 30.
By considering all the probable members , 
the reddening turns out to be E$(B-V)$~=~0.58$\pm$0.05 .\\
This value is in good agreement with previous
estimates. In fact
Moffat \& Vogt (1975) found E$(B-V)$~=~0.54$\pm$0.13, whereas
Fitzgerald \& Mehta (1987) derived E$(B-V)$~=~0.588$\pm$0.032.

\subsection{Age and distance}
Age and distance have been inferred by fitting 
the reddening corrected CMDs with solar metallicity
isochrones (Girardi et al 2000).
This is shown in Fig.~15, where  we plotted all the candidate
members according to reddening. Some stars which lie apart from
the MS are indicated with open circles and considered
probable non members reducing at 24 the number of cluster members.\\
In this figure
the dashed line is an
isochrone for the age of $8 \times 10^{6}$ yrs and
the dotted one an
isochrone for the age of $4 \times 10^{6}$ yrs, 
both shifted
by $(m-M)_o$~=~12.70, in fair agreement with previous
estimates. In fact Moffat \& Vogt (1975) suggested 
$(m-M)_o$~=~12.80, whereas Fitzgerald \& Mehta (1987)
derived $(m-M)_o$~=~12.70.
The distance from the Sun is estimated to be  3.5 kpc and
therefore Bochum~11 is lies between
than Bochum~10 and 9.
The comparison with isochrones suggests that Bochum~11
is a  very young cluster with an age less than 
$4 \times 10^{6}$ yrs.\\
\noindent 
The possibility that pre-MS stars are present in Bochum~11
was already suggested  by  Fitzgerald \& Mehta (1987), who
argued that these stars have to be fainter than $V$~=~12.
The position of the stars in Fig.~15 below $V$~=~12
seems to support this hypothesis,  which
demands further investigations by $H_{\alpha}$ 
equivalent width measurements 
or infrared photometry (see for instance
Vallenari et al 1999).\\

\subsection{Astrometric data}
Tycho~2 provided proper motions measurements for four
stars in the region of Bochum~11. Since the sample
is very small, we refrain from any discussion, and simply
report them in Table~7 for the sake of completeness.
We only notice that HD~93632 has very high proper motion components,
which may be due to measurement errors, since it is actually in a
binary system. Looking at our photometric data there is no reason to believe
that this star does not belong to Bochum~11.

\begin{figure}
\centerline{\psfig{file=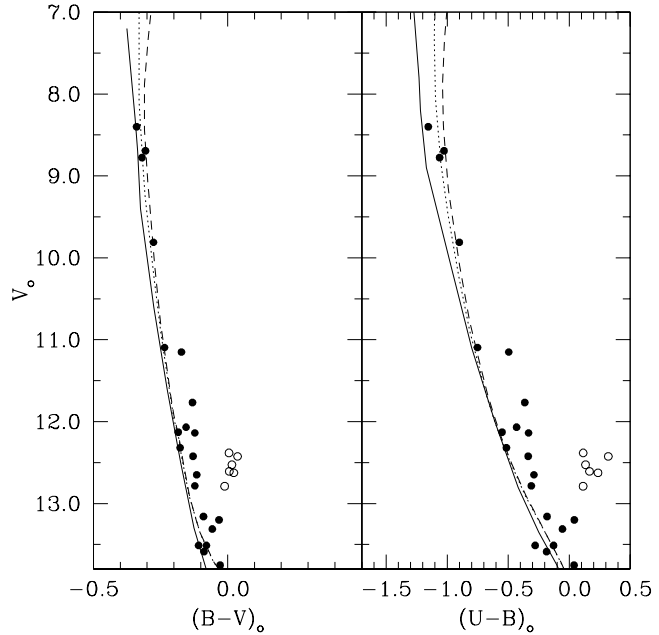,width=9cm,height=9cm}}
\caption{Reddening corrected CMDs for the 
stars in the region of Bochum~11. The solid line is the
empirical ZAMS from Schmidt-Kaler (1982), whereas the dotted
and dashed 
lines are a 4 and 8 $ \times 10^{6}$ yrs isochrones from Girardi et al. (2000).
Open circles indicate cluster non member.}
\end{figure}

\section{Conclusions}
We have presented the first $UBVRI$ CCD photometry for three 
star clusterings in the region of $\eta$~Carinae: Bochum~9,
Bochum~10 and Bochum~11.\\

Bochum~9 is a puzzling object. It seems a group of 30 stars
sharing common reddening located beyond the Carina spiral arm,
although the projected spatial distribution and the poor kinematical
data at disposal seem to imply that we are simply looking at a field
star population.\\

Our analysis suggests that Bochum~10 is a 
very young and poorly populated open cluster. As many other
young poor clusters,  it is certainly an unbounded objects and
will gradually disrupt. We provide estimates of interstellar reddening
and distance compatible with previous studies. \\

As for Bochum~11, we find that it is a young open cluster, less than 
$4 \times 10^{6}$ yrs old, and confirm previous estimates for
the cluster mean reddening and distance.\\

From our photometry we find indications of  possible pre MS
candidates in Bochum~9 and 11. This issue will be addressed
in a forthcoming paper (Romaniello et al 2001), where
$UBVRI$ photometry for all the other star clusterings known
to lie in the Carina spiral feature will be presented 
and compared with theoretical models.

\section*{Acknowledgments}
This paper was based on observations made at ESO-La Silla.
We acknowledge useful discussions with M. Zoccali, 
M. Rejkuba and A. Brown. GC thanks ESO for the kind hospitality.
We are grateful to Drs. A. Feinstein and A.F.J. Moffat for giving
us informations about the instrument set-up used to obtain their
photoelectric photometry.  The referee, Dr. J-C. Mermilliod,
is deeply  acknowledged for important suggestions, which
led to improve the quality of the paper. Finally,
this paper made use of Simbad and WEBDA.

\appendix

\section[]{Photometric solution and errors estimate}

\begin{figure}
\centerline{\psfig{file=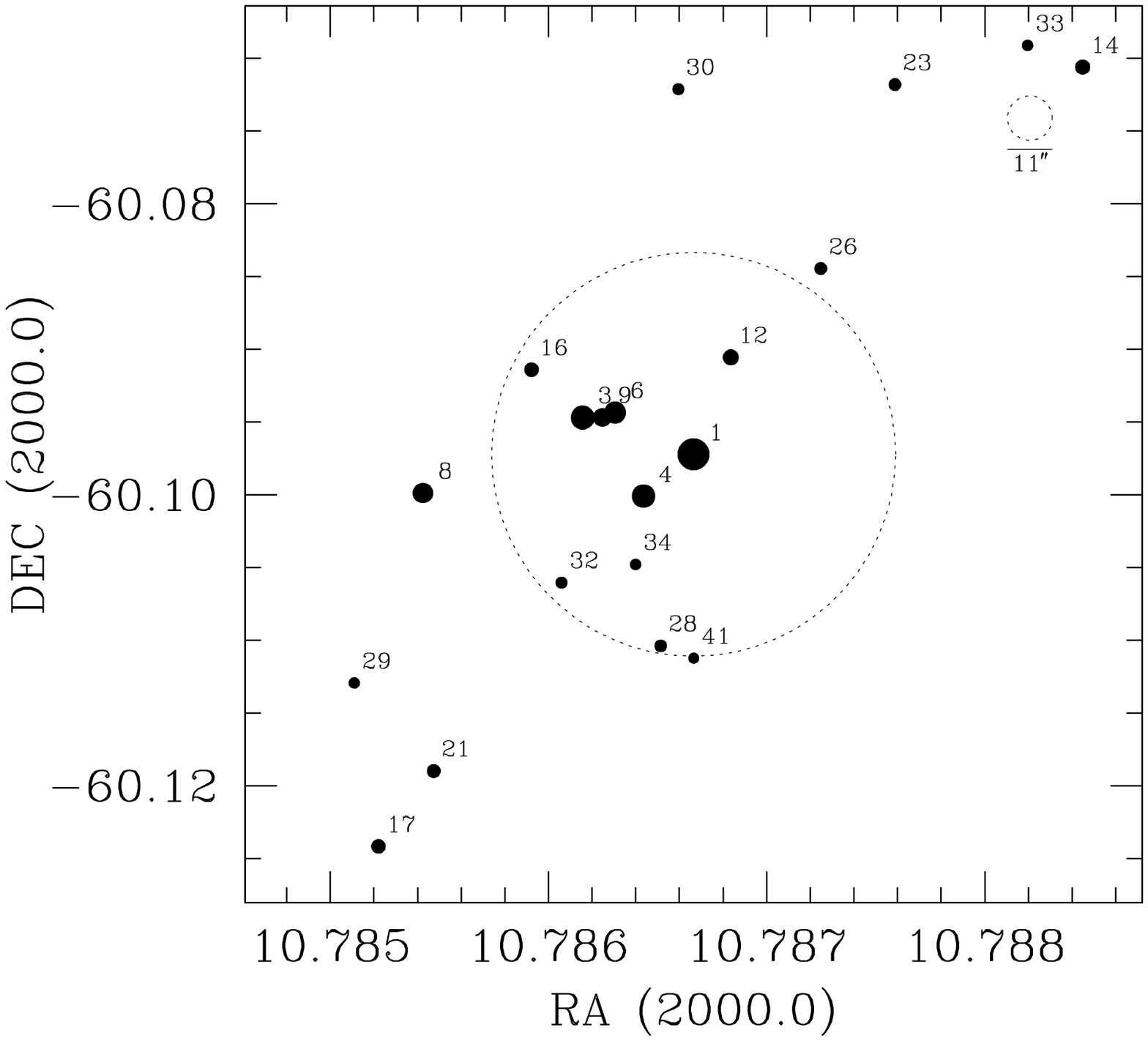,width=8cm,height=8cm}}
\caption{Map of the central region of Bochum~11. The dashed circle 
in the upper right corner indicates the size of
the diaphragm adopted by Moffat \& Vogt (1975), whereas the large
circle ($50^{\prime\prime}$) centered in HD~93632 encloses the cluster core.}
\end{figure}

To allow for a precise photometric calibration we have
observed five multistar fields from the list of Landolt (1992)
each night during the observing run. After bias, flat--field correction and exposure time
normalization, the instrumental magnitudes of the standard stars were
measured using the {\tt qphot} task in {\it IRAF}. To match
Landolt's observations we have used a 14$^{\prime\prime}$ circular
aperture. The same size has been used later to correct DAOPHOT
magnitudes obtained for the scientific targets.

From these measurements we have computed the zero points and the
colour terms of our photometric system via linear least squares
fitting; the results are presented in 
Tab.~\ref{tab:photcoeff}. Since the airmass range covered by our
observations was not large enough to secure a good determination of
the extinction coefficients, we have adopted the average values for
La Silla (cf. Patat 1999), which are also reported in Tab.~\ref{tab:photcoeff}.

\begin{table}
\caption{\label{tab:photcoeff}Average photometric coefficients
obtained during April 13--16, 1996. ESO--Dutch 0.92m telescope, TK CCD~\#33.}
\begin{tabular}{ccccc}
Filter & Ref. Color & $zp$        & $\gamma$ & $k$ \\
\hline  
 $U$   & $(U-B)$    & 19.85$\pm$0.02 & 0.095$\pm$0.020 & 0.46$\pm$0.02 \\
 $B$   & $(B-V)$    & 21.93$\pm$0.01 & 0.079$\pm$0.010 & 0.27$\pm$0.02 \\
 $V$   & $(B-V)$    & 22.19$\pm$0.01 & 0.030$\pm$0.006 & 0.12$\pm$0.02 \\
 $R$   & $(V-R)$    & 22.18$\pm$0.01 & 0.025$\pm$0.014 & 0.09$\pm$0.02 \\
 $I$   & $(V-I)$    & 21.11$\pm$0.01 & 0.062$\pm$0.006 & 0.06$\pm$0.02 \\
\hline
\end{tabular}
\end{table}

\begin{figure*}
\centerline{\psfig{file=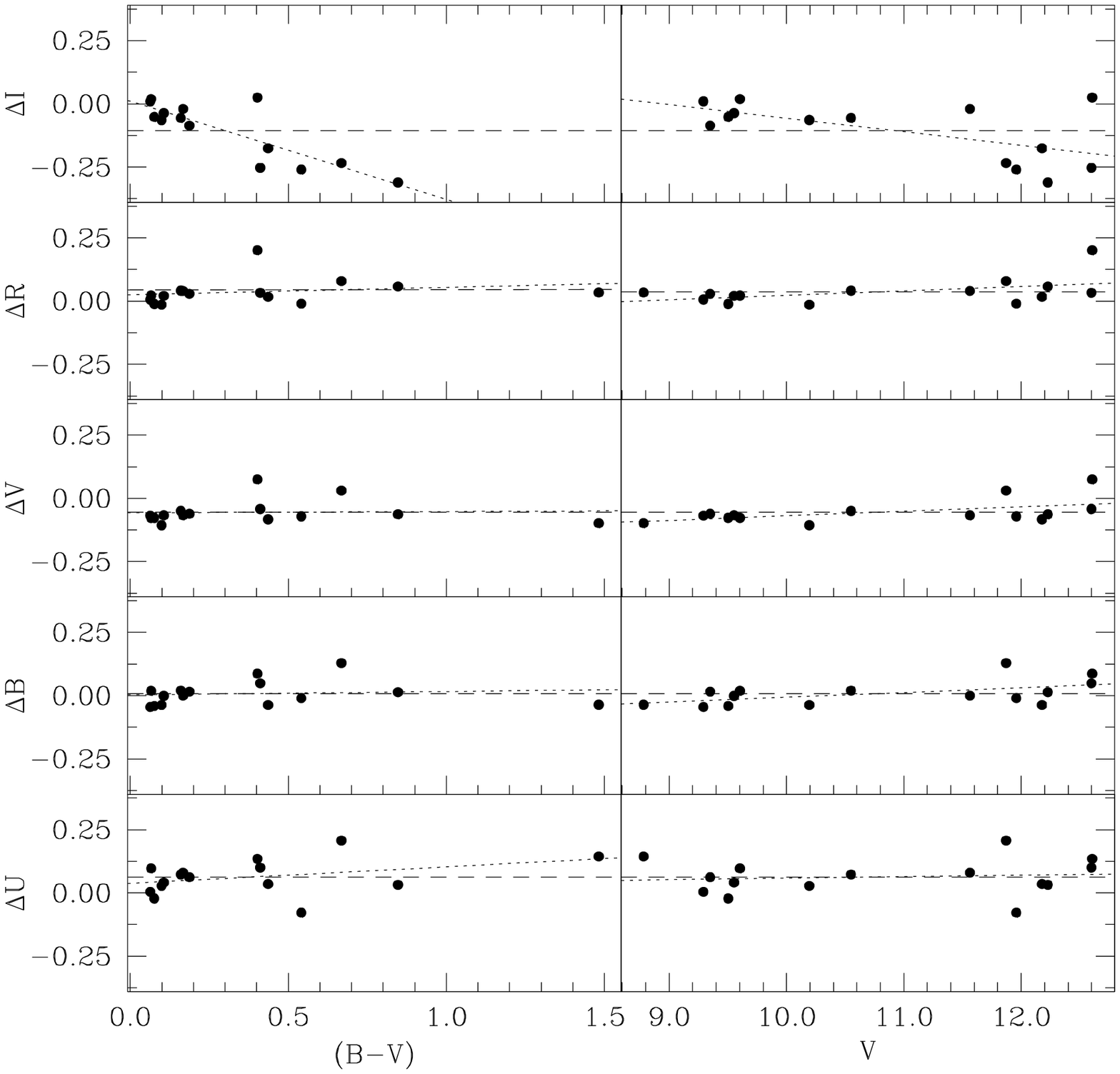,width=16cm,height=18cm}}
\caption{Deviations of photometric measurements presented in this
work from the results published by Feinstein (1981) for
Bochum~10. Dotted lines represent a linear least squares fitting to 
the points, while the  dashed lines indicate the average deviation.}
\label{pippo}
\end{figure*}

The transformation from instrumental magnitudes to the standard
Kron-Cousins system was obtained with expressions of the form

\begin{equation}\label{eq:mag}
M_i=m_i + zp_i + \gamma_i (M_i-M_j) - k_i z
\end{equation}

where $M_i$, $m_i$, $zp_i$, $\gamma_i$ and $k_i$ are the
calibrated magnitude, instrumental magnitude, zero point, colour term
and extinction coefficient for the $i-th$ passband and $z$ is the
airmass. The transformation requires of course the knowledge of the
reference colour $(M_i-M_j)$, which is easily computed from the 
instrumental magnitudes through the following relation:

\begin{equation}\label{eq:col}
(M_i-M_j)=\frac{m_i-m_j + zp_i -zp_j - (k_i-k_j)z}{\gamma_{ij}}
\end{equation}

where we have set $\gamma_{ij}=1-\gamma_i+\gamma_j$. If $\sigma_{mi}$,
$\sigma_{zpi}$, $\sigma_{\gamma i}$ and $\sigma_{ki}$ are the RMS
errors on the instrumental magnitude, zero point, colour term and
extinction coefficient for the $i-th$ passband, formal 
uncertainties on calibrated colors are then obtained propagating
the various errors through Eq.~\ref{eq:col} as follows:

\begin{equation}
\sigma_{(Mi-Mj)}^2 \simeq 
\frac{\sigma_{m,ij}^2 + \sigma_{ps,ij}^2 + (M_i-M_j)^2\sigma_{\gamma,ij}^2}{\gamma_{ij}^2}
\end{equation}

For sake of simplicity, we have set
$\sigma_{m,ij}^2=\sigma_{mi}^2+\sigma_{mj}^2$,
$\sigma_{\gamma,ij}^2=\sigma_{\gamma i}^2+\sigma_{\gamma j}^2$ and
$\sigma_{ps,ij}^2 = \sigma_{zp,ij}^2+z^2\sigma_{k,ij}^2$.

Finally, the RMS uncertainties on the calibrated magnitudes are given by:

\begin{equation}
\sigma_{Mi}^2 \simeq
\sigma_{mi}^2 + \sigma_{psi}^2 +  
(M_i-M_j)^2\sigma_{\gamma i}^2 + \gamma_i^2 \sigma_{(Mi-Mj)}^2
\end{equation}

where we have neglected the error on $z$ and assumed that the
images in different passbands have been obtained at very similar 
airmass, as it was in fact the case.

Estimated uncertainties as a function of magnitude are reported in
Tab.~\ref{tab:photerr}, from which it appears clearly that down to
$V\simeq17$ they are dominated by the errors on the photometric
solution, while at fainter magnitudes the contribution by the
poissonian photon shot noise $\sigma_m$ (estimated by DAOPHOT) 
becomes relevant.

\begin{table}
\caption{\label{tab:photerr} Global photometric RMS errors as a function
of magnitude.}
\centerline{
\begin{tabular}{cccccc}
\hline
Mag    &  $\sigma_U$ & $\sigma_B$ & $\sigma_V$ & $\sigma_R$ & $\sigma_I$ \\
\hline 
  9--11 &  0.04 & 0.03 & 0.03 & 0.03 & 0.03 \\
 11--13 &  0.04 & 0.03 & 0.03 & 0.03 & 0.03\\
 13--15 &  0.04 & 0.03 & 0.03 & 0.03 & 0.03\\
 15--17 &  0.05 & 0.03 & 0.03 & 0.03 & 0.03\\
 17--19 &  0.08 & 0.03 & 0.04 & 0.04 & 0.05\\
 19--20 &    -  & 0.04 & 0.05 & 0.07 & 0.09\\
 20--21 &    -  & 0.07 & 0.09 & 0.15 & 0.22\\
 21--22 &    -  & 0.12 & 0.18 & 0.27 &   - \\
\hline
\end{tabular}
}
\end{table}

As we have already discussed in Secs.~\ref{sec:bo9}, \ref{sec:bo10} and 
\ref{sec:bo11}, the comparison with the data published by other authors
has shown that some deviations exist for the stars which are 
common to the different data sets. Discrepancies on single stars are probably
due to the relatively large diaphragm coupled with the photoelectric
photometer (11$^{\prime\prime}$-14$^{\prime\prime}$, Moffat and Feinstein, private communications).
This is the case, for instance, for the star pairs \#3-\#6 and \#6-\#9
of Bochum~11, as it is shown in Fig.~A1.
One other example can be found in Carraro et al (2001) for NGC~3324.

On the other hand systematic deviations seem to exist. To investigate 
this problem we have analyzed the magnitude deviations as a function
of magnitude and color for Bochum~10, for which a comparison with
published data is possible in all pass-bands. The results are
presented in Fig.\ref{pippo}, were the differences between our measurements
and the results by Feinstein (1981) have been plotted as a function of $(B-V)$
(left panel) and $V$ magnitude (right panel). With the only exception
of the $I$ filter, all pass-bands show small systematic deviations
which are consistent with the estimated photometric errors. For $I$ 
filter a color dependency seems to be there, and this would indicate 
the presence of an unaccounted colour term in the photoelectric
measurements.

\bsp

\label{lastpage}

\end{document}